\documentclass[%
superscriptaddress,
reprint,
amsmath,amssymb,
aps,
prx,
10pt,
twocolumn,
floats,
]{revtex4-2}

\usepackage[greek,english]{babel}
\usepackage{graphicx}
\usepackage{dcolumn}
\usepackage{amsmath}
\usepackage{epsfig}
\usepackage{bm}
\usepackage{amssymb}
\usepackage{natbib}
\usepackage{color}
\usepackage{soul}
\usepackage{hyperref}
\hypersetup{colorlinks=true,linkcolor=blue,citecolor=blue,urlcolor=black}

\usepackage{units}

\begin{document}

\newcommand{\change}[1]{\color{blue} #1 \color{black}}
\newcommand{\comment}[1]{\color{blue}\it #1\color{black} \rm}
\newcommand{\changes}[1]{\color{red} #1 \color{black}}
\newcommand{\seb}[1]{\color{green} \emph{#1 } \color{black}}
\newcommand{\michel}[1]{\color{blue} #1  \color{black}}
\newcommand{\ket}[1]{|#1\rangle}
\newcommand{\ketu}[1]{|\!\!\uparrow\rangle}
\newcommand{\ketd}[1]{|\!\!\downarrow\rangle}
\newcommand{\bra}[1]{\langle #1|}
\newcommand{\half}{\frac{1}{2}}
\newcommand{\roottwo}{\sqrt{2}}
\newcommand{\rf}{\mathrm{rf}}
\newcommand{\Th}{\mathrm{th}}
\newcommand{\Vpcm}[1]{\unit[#1]{V/cm}}
\newcommand{\mVpcm}[1]{\unit[#1]{mV/cm}}
\newcommand{\MHz}[1]{\unit[#1]{MHz}}
\newcommand{\ns}[1]{\unit[#1]{ns}}
\newcommand{\wrf}{\omega_\rf/2\pi}
\newcommand{\mmu}{\ensuremath{\mu}}
\newcommand{\lvlc}{\ensuremath{52c}}
\newcommand{\mtwop}{\ensuremath{51g,m=2}}
\newcommand{\mtwo}{\ensuremath{52f,m=2}}
\newcommand{\lvlmthree}{\ensuremath{m_3}}
\newcommand{\lvlmi}{\ensuremath{m_i}}
\newcommand{\tj}[6]{ \begin{pmatrix}
  #1 & #2 & #3 \\
  #4 & #5 & #6 
\end{pmatrix}}

\newcommand{\Methods}{Appendix}

\title{Measuring the Sr$^+$ $5s_{1/2}$ Land\'e $g$-factor Using Singlet-Triplet Oscillations in a Circular Rydberg State of Strontium}

\author{B. Muraz}
\thanks{These authors contributed equally to this work.}
\affiliation{Laboratoire Kastler Brossel, Coll\`ege de France,
 CNRS, ENS-Universit\'e PSL,
 Sorbonne Universit\'e, \\11, place Marcelin Berthelot, 75005 Paris, France}

\author{M. Pepin}
\thanks{These authors contributed equally to this work.}
\affiliation{Laboratoire Kastler Brossel, Coll\`ege de France,
 CNRS, ENS-Universit\'e PSL,
 Sorbonne Universit\'e, \\11, place Marcelin Berthelot, 75005 Paris, France}

 \author{C. Guimard}
\affiliation{Laboratoire Kastler Brossel, Coll\`ege de France,
 CNRS, ENS-Universit\'e PSL,
 Sorbonne Universit\'e, \\11, place Marcelin Berthelot, 75005 Paris, France}

\author{M. Brune}
\affiliation{Laboratoire Kastler Brossel, Coll\`ege de France,
 CNRS, ENS-Universit\'e PSL,
 Sorbonne Universit\'e, \\11, place Marcelin Berthelot, 75005 Paris, France}

 \author{B. Bakkali-Hassani}
\affiliation{Laboratoire Kastler Brossel, Coll\`ege de France,
 CNRS, ENS-Universit\'e PSL,
 Sorbonne Universit\'e, \\11, place Marcelin Berthelot, 75005 Paris, France}
 
  \author{S. Gleyzes}
 \affiliation{Laboratoire Kastler Brossel, Coll\`ege de France,
 CNRS, ENS-Universit\'e PSL,
 Sorbonne Universit\'e, \\11, place Marcelin Berthelot, 75005 Paris, France}

\hyphenation{Ryd-berg sen-sing ma-ni-fold}

\begin{abstract}
Circular states of strontium are promising platforms for quantum technologies. They exhibit much longer lifetimes than laser-accessible Rydberg states and possess an optically active ionic core that can be manipulated with laser light. 
Here we trap circular states in optical tweezers by exploiting the dynamical polarizability of the ionic core. We measure the decay of the orbital component of a $n=51$ circular state and observe singlet-triplet spin oscillations arising from the interplay of vector light shifts induced by the optical tweezers, spin-orbit coupling associated with the Rydberg electron, and the difference in Land\'e $g$-factor between the ionic core and the Rydberg electrons. By observing the two electron spin dynamics, we measure the spin-orbit interaction and the correction $\delta g \sim 10^{-5}$ of the Land\'e $g$-factor of the Sr$^+$ $5s_{1/2}$ ionic core electron with respect to that of a free electron. This provides a first high-precision measurement of $g_{\textrm{Sr}^+, 5s_{1/2}} = 2.002290(1)$.
\end{abstract}

\date{\today}

\maketitle

\begin{figure}
\centering
\includegraphics[width=\linewidth]{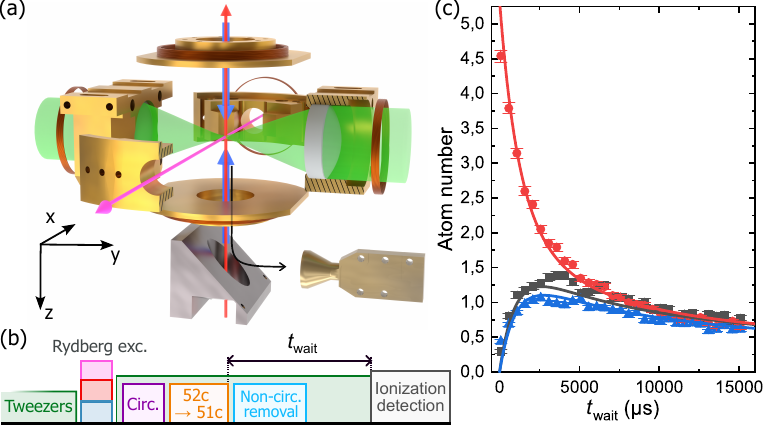}   
\caption{(a) Scheme of the experimental setup. Six gilded electrodes control the electric field at the position of the atoms. For clarity, only half of the front and right electrodes are shown.  The electrodes along the $y$ axis include ITO-coated lenses ($\mathrm{NA} = 0.28$), one of which focuses an incoming 532 nm laser beam (green) into an array of 49 gaussian tweezers (900-$\mu$m waist). Rydberg excitation is achieved using three laser beams at wavelengths of 461 nm (vertical, blue), 767 nm (vertical, red) and 896 nm (horizontal, pink). A $\sigma^-$ polarized rf field at 250 MHz is generated via the electrodes in the horizontal plane. Rydberg atoms are ionized by applying a large electric field across the top and bottom electrodes, and the resulting Sr$^+$ ions are accelerated and deflected (grey electrode) toward a channeltron. Three pairs of coils (orange) control the magnetic field along all directions. (b) Experimental sequence. Tweezers are switched off during the Rydberg excitation. After circularization to $52c$, a microwave pulse prepares $51c$ and a sequence of rf and ionization pulses removes non-circular atoms. After a time $t_\mathrm{wait}$, the atoms are detected by ionization. (c) Time evolution of the circular state populations measured using the channeltron (black: $50c$, red: $51c$, blue: $52c$). The points are experimental with statistical error bars, and the lines correspond to a fit of BBR temperature using a rate-equation model. The populations of the manifolds $n< 52$ are corrected from the variation of detection efficiency of the different $nc$ states (see Appendix \ref{appc}).}
\label{fig1}
\end{figure}

\section{Introduction}

Rydberg atoms are powerful tools for quantum technologies \cite{adams_rydberg_2020}. When prepared in optical tweezer arrays with a few micrometer inter-atomic distance \cite{browaeys_many-body_2020}, the strong dipole-dipole interactions between neighboring atoms enable the emulation of complex quantum dynamics beyond the reach of classical computers \cite{barredo_atom-by-atom_2016,Bernien_Probing_2017,signoles_glassy_2021}. Outstanding results have been obtained in these rapidly evolving domains of quantum simulation and quantum computation \cite{scholl_quantum_2021,ebadi_quantum_2021}. 
Non-interacting Rydberg atoms are also remarkable tools for metrology \cite{fan_atom_2015,Ovsiannikov_rydberg_2011} and atomic physics studies \cite{2017Ramos}.
Their energy structure is linked to fundamental constants \cite{Scheidegger_metrology_2023} and provides insights into the complex structure of multi-electrons atoms \cite{ye_production_2013}. Rydberg atom arrays are also promising for the development of this field of research.

Most Rydberg atom array experiments so far have used laser-accessible, low-angular momentum (``low-$\ell$") states of alkali or alkaline-earth(-like) atoms. Their radiative lifetime is on the order of a few hundred microseconds, limiting the precision of spectroscopic measurements and setting boundaries to the achievable quantum simulation timescales and fidelities.
In contrast, circular Rydberg states \cite{cortinas_laser_2020,teixeira_preparation_2020,Holzl_long-lived_2024}, namely states for which $m_\ell = \pm (n-1)$ \cite{hulet_rydberg_1983}, have much longer lifetimes, in the range of tens of milliseconds at cryogenic temperature, opening up interesting perspectives for quantum simulation \cite{nguyen_towards_2018,Cohen_quantum_2021} and high-resolution spectroscopy of atomic properties \cite{2017Ramos}.

In this context, alkaline-earth and alkaline-earth-like elements are particularly interesting. Once one of the two valence electrons is excited to a circular state, the atom still possesses an optically active ionic core. The ionic core electron can be used for laser cooling \cite{lachaud_slowing_2024} or optical trapping \cite{anderson_trapping_2011,wilson_trapping_2022,Holzl_long-lived_2024}. It can probe the state of the Rydberg electron \cite{pham_coherent_2022, muni_optical_2022, 2024Wirth}. Conversely, we show here that the Rydberg electron provides insights into the properties of the ionic core electron, in particular its Land\'e $g$-factor.

For alkaline-earth atoms, the energy level structure must take into account the spins of both valence electrons. For low-$\ell$ Rydberg states ($\ell \le 3$), the exchange energy lifts the degeneracy between singlet and triplet states \cite{snow_indirect_2003}. For circular states, this interaction is negligible and the spin energy spectrum is only determined by the interaction between each spin and the magnetic field acting on it. If these interactions are different for the core electron and the Rydberg electron, the degeneracy between the states $\ketu \ _c\ketd\ _R$ and $ \ketd \ _c\ketu\ _R$ is lifted, where $|\cdot\rangle_c$ and $|\cdot\rangle_R$ denote the spin state of the ionic core and Rydberg electrons respectively. 
Notably, the spin-orbit interaction results in two different magnetic fields experienced by each of the electrons.
Additionally, in the presence of an external magnetic field, the Land\'e $g$-factor difference, $\delta g$, between the ionic core and the Rydberg electrons also contributes to this splitting. 
When preparing a singlet spin state, one thus expects singlet-triplet oscillations that we used for measuring this energy difference, providing access to both the spin-orbit coupling and the value of $\delta g$.

In this paper we prepare circular states of strontium from individual ground-state atoms trapped in optical tweezers inside a 4.2 K cryostat. First, we trap the circular states in the same optical tweezers as the ground state atoms thanks to the polarizability of the ionic core. We measure a 1.8(1) ms lifetime of the circular states, from which we infer a blackbody radiation (BBR) temperature of 22(1) K. This long lifetime allows us to measure the spin-orbit coupling for different values of the principal quantum number $n\sim49-53$ by recording singlet-triplet oscillations over several milliseconds. Performing similar measurements as a function of the external magnetic field, we extract the difference of Land\'e $g$-factor between the ionic core and the Rydberg electrons $\delta g = -2.9(1)\times 10^{-5}$. 

\section{Experimental set-up}

The experimental setup is shown in Fig.1(a). We prepare a broad-line blue magneto-optical trap (MOT) of $^{88}$Sr atoms from a Zeeman slower along the $x$ axis (not shown). We transfer the atom to a narrow-line red MOT in order to load an array of $7 \times 7$ optical tweezers, separated by 30\,\textgreek{m}m, created from a 532\,nm wavelength trapping laser with linear polarization along the $z$ axis. We prepare and trap ground state atoms using 2.7 mW per trap corresponding to a trap depth of 0.5 mK. We then apply a chirp-cooling procedure to simultaneously cool the atoms and perform parity projection, leading to either zero or one atom in each trap \cite{2023Holzl} (see Appendix \ref{appa}). We finally lower adiabatically the tweezer power from 2.7 to 0.4 mW and obtain about $\sim 25$ trapped single atoms in their electronic ground state, at a temperature of 3\,\textgreek{m}K, randomly distributed across the trap array.

To prepare the circular states, we briefly switch off the trapping laser and transfer the atoms from the $5s^2\,^1S_0$ into the $5s 52f\,^1F_3$ level through a resonant three-photon excitation (461 nm, 767 nm, and 896 nm) via the $5s5p\,^1P_1$ and $5s5d\,^1D_2$ intermediate levels \cite{teixeira_preparation_2020}. The laser excitation is performed at zero electric field and in a magnetic field $\mathbf{B}_0 ={B}_0\hat{\boldsymbol{z}}$, with $|{B}_0|$ ranging from 0.27\,G to 6.28\,G, which lifts the degeneracy between the magnetic sublevels $m_\ell$ of the final state. The excitation lasers are tuned to selectively transfer the atoms into the $m_\ell=-3$ state. We then switch on the trapping laser with a different power per tweezer $P_\mathrm{Ryd}$. We apply a 2\,\textgreek{m}s resonant 461~nm laser pulse to expel atoms that remained in the $5s^2\,^1S_0$ state due to the limited Rydberg excitation efficiency ($\sim 35~\%$), leaving only Rydberg atoms in the traps. 
Next, we ramp the electric field $\mathbf F =F\hat{ \mathbf z}$  to $F = 2.22(2)$\,V/cm in 1 \textgreek{m}s 
and we adiabatically transfer the atoms to the circular state $52c$ with $m_\ell=-51$ by increasing the electric field to $2.63(2)$\,V/cm in the presence of a $\sigma_-$ polarized radio-frequency field at 250 MHz \cite{teixeira_preparation_2020}. This prepares 
the atom in the $5s\,52c$ singlet spin state (see Appendix \ref{appc}).
We then maintain an electric field of $2.63(2)$\,V/cm to protect this state from mixing with other levels in the $n$ manifold. Finally, we detect the Rydberg atoms by ramping the electric field up to 163(2) V/cm to ionize them and accelerate the produced Sr$^+$ ions onto a channeltron. As the ionization threshold depends on the Rydberg state, the distribution of ion arrival times allows us to reconstruct the relative populations in the different manifolds (see Appendix \ref{appc}).

The preparation fidelity is limited, and only 66(2)\% of the Rydberg atoms end up in the $52c$ state. To prepare pure circular atoms, we first selectively transfer the $52c$ atoms into the $51c$ state using a microwave pulse. A short radio-frequency pulse sequence then drives the non-circular atoms left in the $n=52$ manifold back to low-angular-momentum states while leaving the $51c$ states unaffected. 
Finally, we apply an electric field pulse that only ionizes the low-angular-momentum states, ensuring that the optical tweezers only contain atoms in $51c$ (see Appendix \ref{appc}).

\section{Lifetime measurement}

We first measure the lifetime of the $51c$ circular state. Spontaneous emission and thermal transfer mainly induce transitions between adjacent circular states $nc \rightarrow (n\pm1)c$ \cite{2023Wu, cantat-moltrecht_long-lived_2020}. We measure the evolution of the circular state populations $52c$, $51c$ and $50c$ as a function of the time delay $t_\mathrm{wait}$ between the $52c\rightarrow51c$ microwave pulse and the start of the ionization ramp (Fig.\,\ref{fig1}.c)  
with $B_0 = 3.160(1)$\,G and $P_\mathrm{Ryd}=1.35(4)$ mW.
The data are in good agreement  
with a fit to a rate-equation model, from which we extract a BBR temperature of 22(1)\,K. 
This temperature is higher than that of the cryostat. This difference could be reduced by improving the BBR thermalization using cold microwave absorber. At this temperature, the lifetime for the $51c$ circular state is 1.8(1)\,ms, already $\sim$40 times longer than that of the initially prepared $5s 52f\,^1F_3$.

\begin{figure}
\centering
\includegraphics[width= \linewidth]{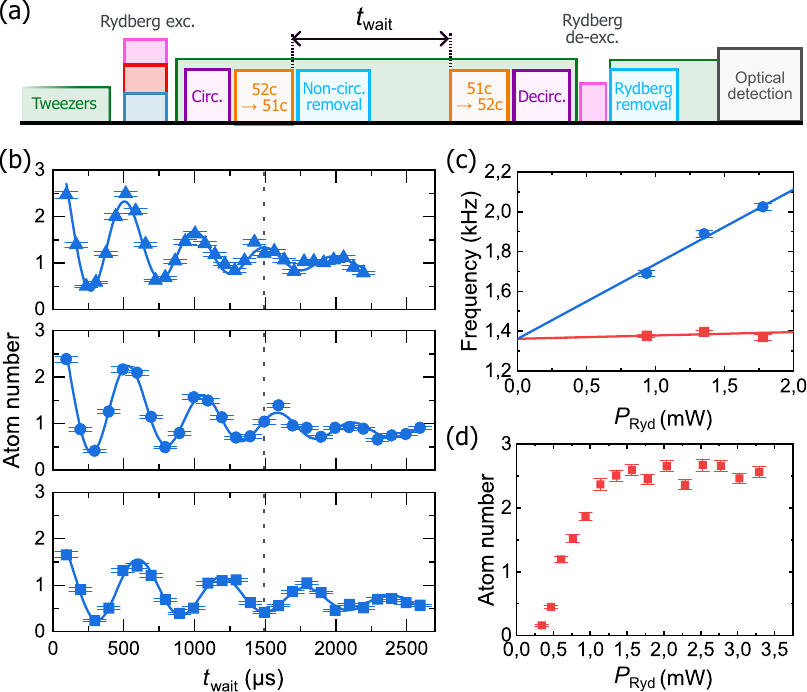}
\caption{(a) Experimental sequence. The sequence is identical to that of Fig\, \ref{fig1}(a) until the end of $t_\mathrm{wait}$, after which we apply a decircularization sequence (time-reversal of the circularization sequence). After decircularization, the tweezers are briefly switched off and the atoms are transferred back to the ground state using a short 896 nm laser pulse, before a large electric field ionizes the remaining Rydberg atoms. (b) Averaged number of atoms detected over all traps as a function of $t_\mathrm{wait}$ for different tweezer powers $P_\mathrm{Ryd}$ [squares: 0.94(3) mW, disks: 1.35(4) mW, triangles: 1.77(5) mW]. 
The solid lines are fits to the expression $N(t) = e^{-t/t_1}[Ae^{-t/t_2}\sin({2\pi \nu t +\phi_0) }+y_1]$ where $t_1,A,t_2,\nu ,\phi_0,y_1$ are free parameters. 
The dotted line is a guide to the eye indicating the position of the third minimum for the smallest trap depth. (c) Variation of the singlet-triplet oscillation frequency as a function of the tweezer power with (red) and without (blue) the compensating field $B_y$.  
The points are extracted from the fits to the oscillations, and the lines are linear fits constrained to share the same value at zero intensity. (d) Number of recaptured atoms for a fixed value of  $ t_\mathrm{wait}=94.3$ \textgreek{m}s as a function of the tweezer power $P_\mathrm{Ryd}$. The points are experimental with statistical error bars. }
\label{fig2}
\end{figure}

The ionization detection method has a large detection area without spatial selectivity and thus does not directly show that the atoms are trapped. It does not discriminate singlet from triplet states either.   We thus devise an optical detection method which overcomes these two limitations by selectively detecting the trapped $51c$ singlet atoms.
The experimental sequence is depicted in figure~\ref{fig2}(a). 
Instead of ionizing the atom, we transfer back the $51c$ atoms into $52c$ with a microwave pulse, perform the time reversal of the adiabatic passage and ramp the electric field down to zero.
 The singlet circular atoms are transferred back to the singlet $5s 52f\,^1F_3$ state, while circular atoms in the triplet state end up in one of the $5s 52f\,^3F_J$ levels. We then switch off the trapping laser for 4 \textgreek{m}s during which we apply a short 896 nm laser pulse. The $5s 52f\,^1F_3$ atoms are transferred to the $5s4d\,^1D_2$ state, from which they rapidly decay to the $5s^2\,^1S_0$ ground state. The $5s 52f\,^3F_J$ states, however, are not affected by the 896 nm light, since for low-angular-momentum states the exchange energy lifts the degeneracy between singlet and triplet states. The trapping light is then switched back on, at the power $P_\mathrm{Ryd}$, and the ground-state atoms are recaptured in the optical tweezers. We subsequently apply a large electric field to ionize the remaining Rydberg atoms. This step prevents the triplet $F$ states from decaying back to the ground state by spontaneous emission.
The atoms are imaged 20 ms later using fast single atom imaging \cite{2025Su}. We have checked that this time is long enough for the untraped atoms to have flown away from the imaging region, thus ensuring that only the atoms that were recaptured contribute to the atomic count deduced from the image. The number of recaptured atoms is proportional to the population of the $51c$ singlet state at the end of $t_\mathrm{wait}$.

\section{Singlet-triplet oscillations}
\label{seq-ST}

We now use this optical detection method to observe the evolution of the $51c$ singlet state population.
Figure \ref{fig2}(b) presents the average atom number detected in the optical tweezers as a function of the time delay $t_\mathrm{wait}$ [timing in Fig.\,\ref{fig2}(a)] for different tweezer powers $P_\mathrm{Ryd}$. We observe damped oscillations of the recapture signal as a function of $t_\mathrm{wait}$. The damping time is qualitatively similar to the decay expected from BBR, demonstrating that the trapping is efficient on the timescale of the radiative lifetime. 

The oscillations are the signature of the singlet-triplet dynamics. 
We first observe that the frequency of the singlet-triplet oscillations depends on the tweezer power [blue points in Fig\,\ref{fig2}(c)]. 
This may originate from imperfect polarization of the incoming beam or birefringence induced by the stress on the cold windows inside the cryostat. 
If the trapping light is not perfectly linearly polarized, it induces a differential light shift between the two different $m_{j_c}$ ionic core sublevels resulting in an effective magnetic field $\mathbf B_\mathrm{eff}$, acting on the ionic core electron only, oriented along the beam propagation axis and proportional to the trapping laser intensity.
In the limit $|| \mathbf B_0 ||\gg ||\mathbf B_\mathrm{eff}||$, the effective field $\mathbf B_\mathrm{eff}$, perpendicular to the directing field $\mathbf B_0$, should not affect to first order the spin oscillation frequency.
A small angle of $\sim0.2^\circ$ between $\mathbf B_\mathrm{eff}$  and  $\hat{ \mathbf y}$, well within the geometrical uncertainty of the experimental set-up, is sufficient to explain the observed variation of the spin oscillation frequency with the trapping power.
We cancel to first order this dependency by adding a small magnetic field $B_y=-0.083(3)$~G along $\hat{ \mathbf y}$ to make the total magnetic field perpendicular to $ \mathbf B_\mathrm{eff}$ [red points in Fig\,\ref{fig2}(c)].

In these conditions, we measure the number of recaptured atoms for a fixed value of  $ t_\mathrm{wait}=94.3$ \textgreek{m}s  for different values of the power $P_\mathrm{Ryd}$ [Fig. \ref{fig2}(d)].
In the absence of trap ($P_\mathrm{Ryd}=0$) we detect no atom in the final image. As the trapping laser intensity increases, the recaptured atom number increases gradually. For a power larger than 1.35 mW per tweezer, corresponding to a trap depth of 51~\textgreek{m}K, the number of detected atom saturates, demonstrating that the trapping potential is sufficient to confine the Rydberg atoms.

\section{Spin-orbit coupling and Land\'e factor measurement}
\label{seq-varyn}

\begin{figure}
\centering
\includegraphics[width= \linewidth]{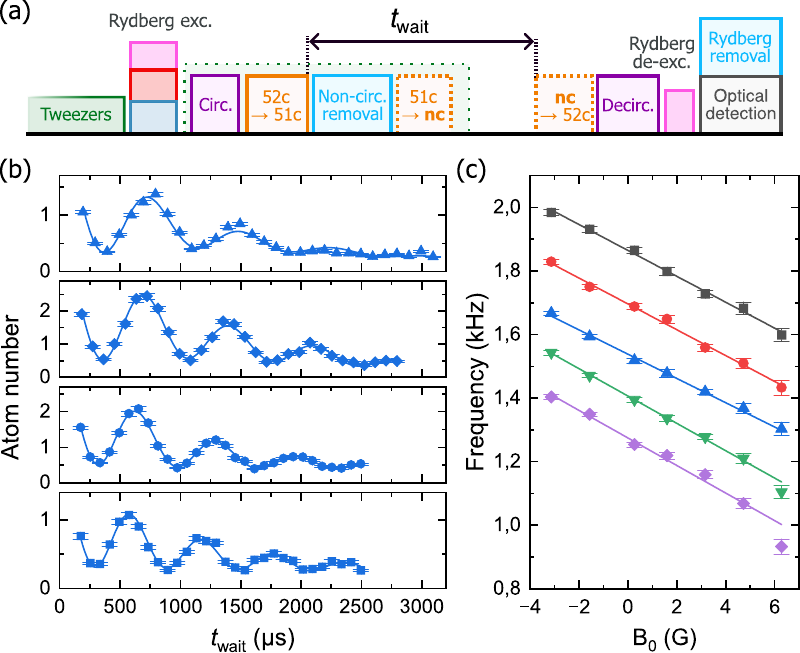}
\caption{(a) Experimental sequence.  The tweezers are not switched back on during $t_\mathrm{wait}$ (except shortly for $B_0<0$, dotted green, see Appendix \ref{appa}). To detect the atoms, we collect their fluorescence immediately after transferring the singlet circular states back to the ground state. (b) Spin oscillations recorded for $n = 51$ without trapping the Rydberg atoms, for different values of $B_0$ [squares: $-3.130(1)$\,G, disks: $0.272(1)$\,G, diamonds: $3.160(1)$\,G, triangles: $6.280(1)$\,G]. The differences in amplitude between the different datasets result from variations in laser excitation efficiency. The points are experimental with statistical error bars, and the solid lines is a fit to the same model as in Fig. 2(b). (c) Spin oscillation frequencies as a function of $B_0$ for different $n$ (black: $n=49$, red: $n=50$, blue: $n=51$, green: $n=52$ and purple: $n=53$). The points are the fit results with statistical error bars, the lines are independent linear fits.}
\label{fig3}
\end{figure}

We now take advantage of the spin selective detection method to precisely measure the spin-orbit coupling as well as the $\delta g$ between the two valence electrons. In order to get rid of the systematic effects due to the trapping laser polarization imperfections, we perform this measurement at $P_\mathrm{Ryd}=0$ and image the atoms immediately after transferring the singlet circular states back into the ground state. When the tweezers are off, the atoms move away from the traps. However, they remain within our field of view , 
long enough (see Appendix \ref{appa}) to observe a few periods of singlet-triplet oscillations [Fig\, \ref{fig3}.(a)].

We first consider the spin-orbit effect on the singlet-triplet oscillations. The circular motion of the Rydberg electron induces a magnetic field $\mathbf B_{\mathrm{so},n}$ at the position of the ionic core. Similarly the Rydberg electron experiences a field $-\mathbf B_{\mathrm{so},n}/2$ due to its rotation around the Sr$^+$ ionic core, the factor 1/2 corresponding to the well-known relativistic Thomas precession factor. The two spin states $\ketu \ _c\ketd\ _R$ and $ \ketd \ _c\ketu\ _R$ are thus separated by an energy
\begin{equation} 
\label{eqn-nuso} 
h\nu_{\mathrm{so}}(n) = \frac 3 2 g_e \mu_B |\mathbf B_{\mathrm{so},n}|,
\end{equation} 
where $\mu_B$ is the Bohr magneton and $g_e$ the free-electron Land\'e $g$-factor. Here we assumed that the $g$-factor of the ionic core $g_c$ and the Rydberg electron $g_R$ to be both equal to $g_e$.  For $n = 51$, we expect $\nu_{\mathrm{so}} = 1.539~$\,kHz (see Appendix \ref{appd}).

In the presence of a magnetic field $B_0\hat{\boldsymbol{z}}$ along the quantization axis,
 the singlet-triplet oscillation frequency also depends on  the deviation of $g_c$ and $g_R$ from $g_e$. For the Rydberg electron, we expect $(g_R-g_e)/g_e \sim 10^{-8}$ for $n\sim 50$ \cite{2006Jentschura}. For the ionic core electron, however, relativistic effects \cite{breit_magnetic_1928} and core polarization \cite{lindroth_ab_1993} lead to larger corrections to the Land\'e factor with $(g_c-g_e)/g_e \sim 10^{-5}$ \cite{tommaseo_mathsfg_scriptscriptstyle_2003}. The two states $ \ketu \ _c\ketd\ _R$ and $ \ketd \ _c\ketu\ _R$ thus experience a differential Zeeman shift $\delta g  \mu_B B_0$, with $\delta g = (g_c - g_R)$. For our choice of circular state with $m_\ell < 0$, the singlet-triplet oscillation frequency becomes 
\begin{equation}
\label{eq:nu_spin}
\nu_{\mathrm{spin}}(n) = \nu_{\mathrm{so}} (n) +  \delta g \frac{\mu_B } h B_0 . 
\end{equation}

We record the spin oscillations for different values of $n$ and $B_0$ [Fig\, \ref{fig3}.(b)] and we deduce the value of $\nu_{\mathrm{spin}}(n)$ from a fit of the data. Figure\, \ref{fig3}.(c) shows that for each value of $n$, $\nu_{\mathrm{spin}}(n)$  has the expected linear variation with $B_0$, consistent with Eq. \ref{eq:nu_spin}.
From a linear fit to the data, we extract $\nu_{\mathrm{so}}(n)$ for $n = 49, \ldots, 53$ [Fig.\,\ref{fig4}(a)]. The values agrees well with \emph{ab initio} calculations (green line, see Appendix \ref{appd}). From the slope of these fits we deduce a value of $\delta g$ for each value of $n$ [Fig.\,\ref{fig4}(b)]. We do not observe a significant dependency with $n$.  A global fit of the same data with the slope as a shared parameter yields $\mu_B\delta g/h = -0.041(1)$ kHz/G, corresponding to $\delta g/g_{e^-} = -1.47(5) \times 10^{-5}$. At this precision, we can assume that the Land\'e $g$-factor of the Rydberg electron is that of a free electron \cite{2006Jentschura}, and therefore our experiment provides a first high-precision measurement of $g_{\textrm{Sr}^+, 5s_{1/2}} = g_e + \delta g = 2.002290(1)$ \cite{ludlow_optical_2015}. To the best of our knowledge, there is no measurement nor theoretical prediction for this $g$-factor. However, our result is consistent with the $g$-factor values measured for other alkali-like ions [Fig.\,\ref{fig4}(c)].

\section{Conclusion and outlook}

We have demonstrated the preparation and trapping of circular Rydberg states in the same gaussian 532 nm wavelength optical tweezers as for trapping the ground state of strontium thanks to the ionic core polarizability in a cryogenic environment. We measure a circular state lifetime  of 1.8(1)\,ms, corresponding to a blackbody radiation temperature of 22(1)\,K. This long lifetime allows us to observe several periods of singlet-triplet oscillations. We identify three effects that contribute to these oscillations. The first contribution comes from the residual polarisation imperfections of the trapping beam that induce a differential light shift on the ionic core electron spin states. The second contribution is due to the spin-orbit interaction. Finally, a small correction to the singlet-triplet oscillation frequency arises from the different gyromagnetic factors of the Rydberg and ionic core electron.

This experiment provides the first measurement of spin-orbit coupling in circular Rydberg states and of the Land\'e $g$-factor of the $5s_{1/2}$ state of the Sr$^+$ ionic core.
The dynamics of the entanglement between the two valence electrons gives direct access to the differential Land\'e $g$-factor, enabling the measurement of  minute corrections in the $10^{-5}$ range directly from the atomic spin oscillation signal, without the need for an extremely precise calibration of the external magnetic field or its gradient \cite{tommaseo_mathsfg_scriptscriptstyle_2003}. This demonstrates the high potentential of of alkaline-earth circular Rydberg atoms for atomic spectroscopy and metrology.

\begin{figure}
\centering
\includegraphics[width=\linewidth]{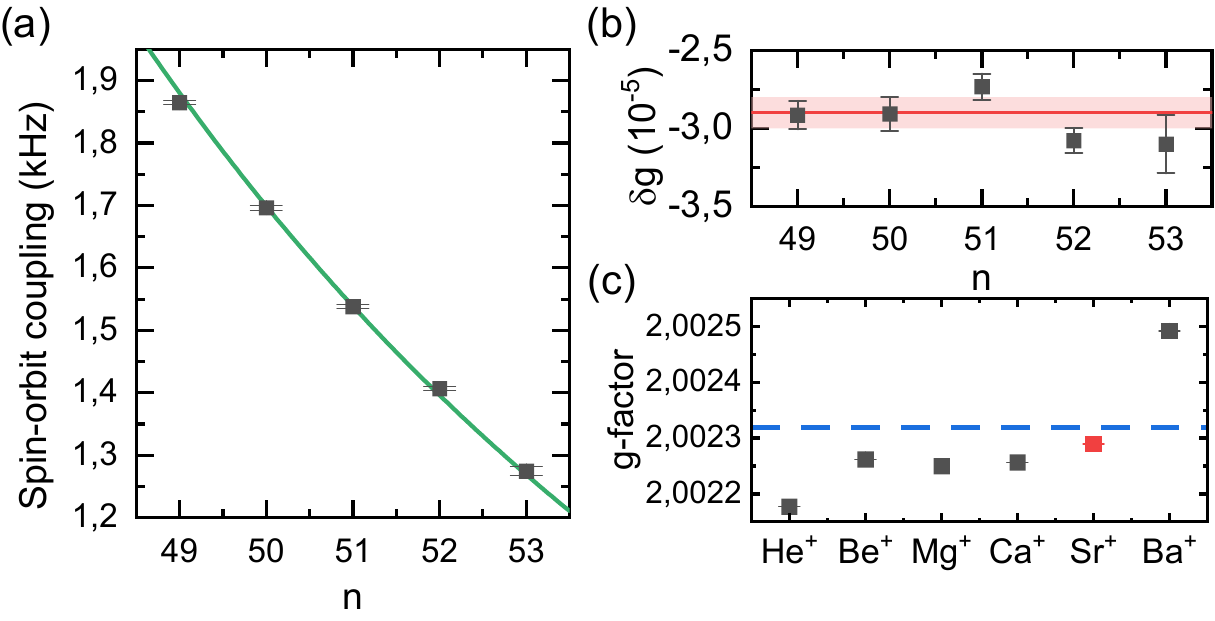}
\caption{(a) Values of $\nu_\mathrm{so}(n)$ for $n$ ranging from 49 to 53. The points correspond to the intercepts of the linear fits in Fig.\,\ref{fig3}(b) with statistical error bars. The solid line is the theoretical value (no free parameter). (b) Measurement of $\delta g$. The points are the slope of each individual fits of Fig.\,\ref{fig3}(b) with statistical error bars, while the solid line shows the result of a global fit of data of Fig.\,\ref{fig3}(b) with $\delta g$ as a shared parameter. The red-shaded area shows the statistical error bar of the global fit. (c) Measured $g$-factors of different alkali-like ions \cite{tommaseo_mathsfg_scriptscriptstyle_2003}, with the free-electron value shown as the blue dashed line. This work is shown in red. }
\label{fig4}
\end{figure}

\begin{acknowledgements}
We thank Y. Machu, I. Dotsenko, P. Méhaignerie for their software development, J. Beugnon for lending us the trapping laser, C. Sayrin, P. Indelicato, J.M. Raimond and R. Ozeri for fruitful discussions. 

This publication has received funding by the France 2030 programs of the French National Research Agency (Grant No. ANR-22-PETQ-0004, project QuBitAF), under Horizon Europe programme HORIZON-CL4-2022-QUANTUM-02-SGA via the project 101113690 (PASQuanS2.1). It has been supported by the Quantum Information Center Sorbonne as part of the program investissements d’excellence – IDEX of the Alliance Sorbonne Université.
This research was funded in part by ANR project ANR-23-CE47-0008-01( SCiRQ). A CC-BY public copyright license has been applied by the authors to the present document and will be applied to all subsequent versions up to the Author Accepted Manuscript arising from this submission, in accordance with the grant’s open access conditions. 

Data are available upon reasonable request.

\end{acknowledgements}

\appendix

\section{Individual atoms in optical tweezers}
\label{appa}

\subsection{Preparation of ground state atoms in optical tweezers}

Atoms effusing from an oven are slowed down using a 9 cm-long Zeeman slower placed inside the cryogenic environment and load during 150 ms a ``blue" broad-line magneto-optical trap (MOT) at a wavelength of 461 nm. 

The optical tweezers are produced by focusing a 532\,nm laser beam reflected from a spatial light modulator (SLM) with an aspheric lens of numerical aperture $\textrm{NA} = 0.28$, using the field aberration correction method described in Ref.\,\cite{2025Machu}. We use an array of $7 \times 7$ traps separated by 30\,\textgreek{m}m. 

The tweezers are switched on at the end of the blue MOT sequence. To load the optical traps, we first transfer the blue MOT atoms into a narrow-line MOT at a wavelength of 689 nm in 35 ms using the SWAP technique. We then apply single-frequency cooling for 15 ms and switch off the MOT. Finally we perform parity projection and laser cooling simultaneously using the chirp-cooling method similar to \cite{2023Holzl}. 

A bias magnetic field is switched on along the $z$-direction, after which we wait during 150 ms for the field to stabilize (see details on the magnetic field calibration in the next section). The tweezer power is subsequently lowered adiabatically from $2.7$ to $0.4$ mW per tweezer.

\subsection{Optical detection}

We employ a fast imaging scheme inspired by Ref.\,\cite{2025Su}. With the trapping light switched off, the atoms are illuminated with two counter-propagating laser beams along the $z$ axis, at resonance with the 461 nm transition of $^{88}$Sr atoms. The light is linearly polarized along the $x$ direction, perpendicular to the aspheric lens axis ($y$ axis), to maximize the photon collection efficiency. To limit spatial diffusion of the atoms, the two imaging beams at 461 nm strongly saturate the 461 nm transition [saturation parameter $s = 7(1)$] and are switched on alternatively for 2\,\textgreek{m}s, for a total imaging duration of $30$\,\textgreek{m}s. The fluorescence signal is collected through the aspheric lenses used for generating the tweezers, and imaged onto an iXon Ultra 888 EMCCD camera (Andor). 

When the Rydberg atoms are trapped (data shown in Fig.\,\ref{fig2}), imaging is performed 20\,ms after their preparation in order to allow all untrapped atoms to leave the imaging area.
Each tweezer image is binarized by thresholding the photon count over a region of interest (ROI) of $15\times15$ \textgreek{m}m$^2$ (with 1 pixel/\textgreek{m}m). This results in a total photon number of $\sim$15 per atom, and yields an imaging fidelity of $96(1)$\%. 

In the case of untrapped atoms (data shown in Fig.\,\ref{fig3}), imaging is performed immediately after de-exciting the atoms from the Rydberg state. To account for the motion of the atoms due to their initial velocity and to forces due to the residual magnetic or electric field gradients during to the time of flight $t_\mathrm{wait}$, the tweezer image are binarized using large $29\times29$ \textgreek{m}m$^2$ adjacent ROIs.

We observe that the atomic motion depends on the sign of $B_0$, the atoms leaving the ROI region faster for $B_0<0$.
For data with $B_0>0$ the tweezers are kept off after the Rydberg excitation. 
For data with $B_0<0$, we maximize the number of detected atoms by switching the tweezers on for 70 \textgreek{m}s after Rydberg excitation.

\subsection{Calibration of trap parameters}

To calibrate the in-situ the tweezers power, we first measure the differential light-shift induced on the $^1S_0 - {^3P_1}, m_j =\pm1$ transition using 689\,nm laser spectroscopy for a total trapping laser power of 59.8(5) mW measured after crossing the cryostat. 
Knowing the dynamical polarizabilities of the $^1S_0$ and $^3P_1,m =\pm1$ states at 532 nm\,\cite{2018Cooper}, we deduce from this measurement the tweezer trap depth for ground-state atoms $U/k_B = 0.27(1)$\,mK. 
We then measure the radial trapping frequency $\nu_r$. We first release the atoms for 10\,\textgreek{m}s to induce breathing dynamics, release them a second time at a variable time for 20\,\textgreek{m}s and measure their recapture probability. We find $\nu_r=48(1)$~kHz. Assuming a gaussian shape of the tweezers, the combined measurements of $\nu_r$ and $U$ provide a determination of the in-situ tweezer power $P=1.44(5)$\,mW and waist $w = 895(10)$\,nm. This allows us to compute the proportionality factor between the in-situ tweezer power and the power of the trapping laser measured after crossing the cryostat.

The atom temperature after chirp-cooling and adiabatic decompression to 0.4 mW per tweezer is determined using a standard release-and-recapture method \cite{2008Tuchendler}, from which we deduce a temperature of 3\,\textgreek{m}K for ground-state atoms.

We estimate the Rydberg atom polarizability by summing the polarizability of the $5s_{1/2}$ ionic core (221.3 a.u.,  \cite{2025Portal}) and that of the Rydberg electron (-136.3 a.u.), assuming the Rydberg electron has the same polarizability as a free electron \cite{nguyen_towards_2018}. We find that the total Rydberg atom polarizability at the  532 nm wavelength is 8.9 times smaller than that of the $^1S_0$ ground state (756.3 a.u.). We use this value to estimate the trap depth of the optical tweezers for the Rydberg atom $U_{\mathrm{Ry}}$ as a function of the 532 nm laser power.

\section{External field calibration}
\label{appb}
\begin{figure}
\centering
\includegraphics[width=\linewidth]{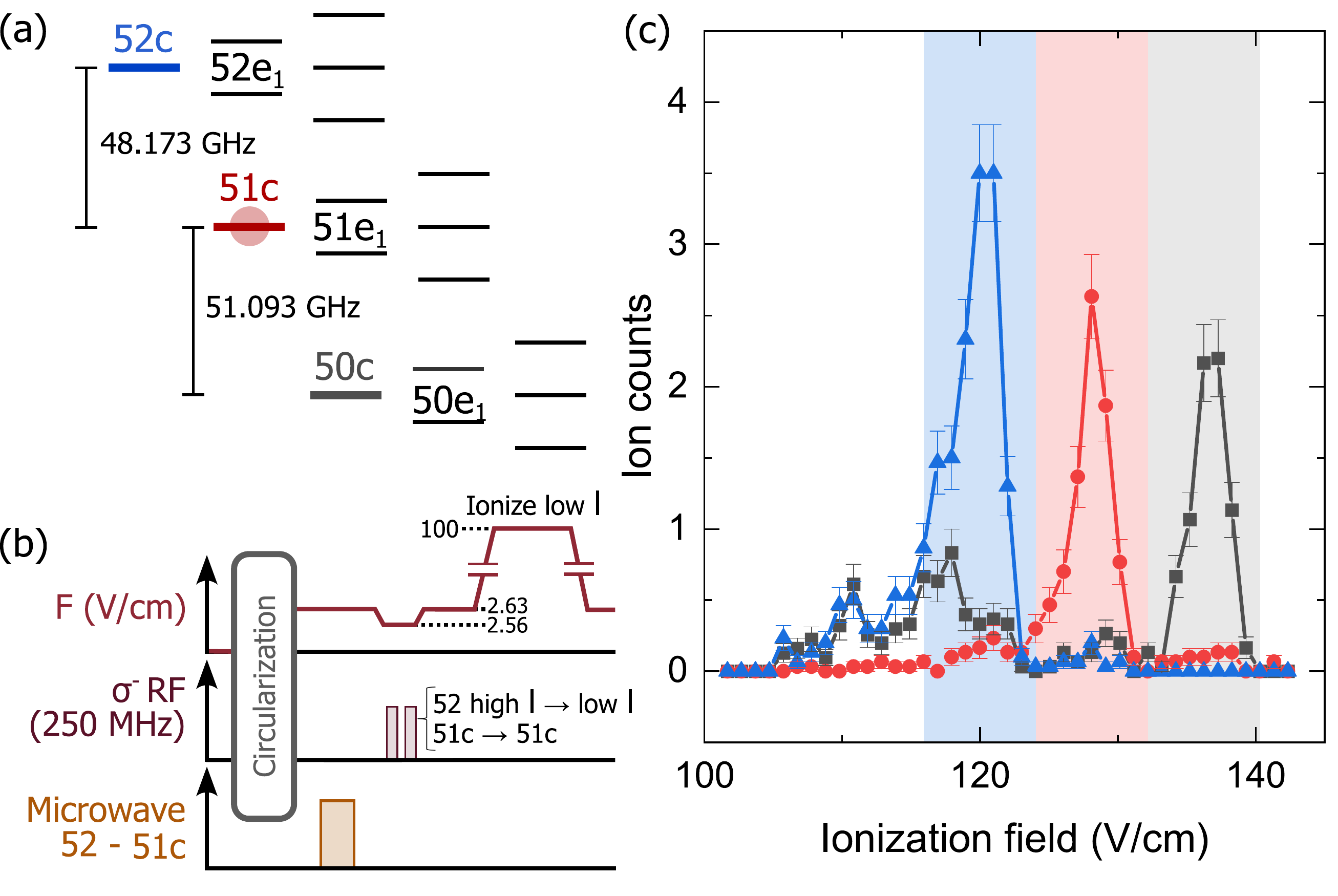}   
\caption{(a) Schematic of the main high-angular momentum Rydberg energy levels involved in the experiment sorted by $m_\ell$. (b) State purification sequence: after selectively transferring the circular atoms from $n=52$ to $n=51$ with a microwave pulse, a sequence of radio-frequency pulses at $250\,\mathrm{MHz}$ brings the remaining high-$\ell$ states in $n=52$ down to low-angular-momentum states. A DC electric field is then ramped to a large value to ionize these low-$\ell$ states, while leaving $51c$ unaffected. (c) Number of ions detected with the channeltron as a function of the ionization field. Each point corresponds to a $20\,\mu\mathrm{s}$ detection window in the arrival time centered around the corresponding ionization field on the horizontal axis. The curves correspond to different experimental sequences: (blue) at the end of circularization towards $52c$, without state purification; (black) after a two-photon microwave transfer from $52c$ to $50c$; (red) after purification, showing a pure $51c$. The colored regions (blue, red, grey) indicate the detection windows used to count the populations of the different circular states ($52c$, $51c$, $50c$, respectively) shown in Fig.\,\ref{fig1}.d.}
\label{figMethods_Channeltron}
\end{figure}

\subsection{Electric field}

The magnitude of the electric field $F$ was calibrated by measuring the difference between the resonant frequencies of the Rydberg transitions $52c \leftrightarrow 50c$ and $52e_1 \leftrightarrow 50e_1$, where $ne_1$ denotes the elliptical state defined in Fig.\,\ref{figMethods_Channeltron}.a. Although each of these two transitions depend linearly on the magnetic field through the Zeeman shift (since $\Delta m = -1$), taking their difference eliminates this dependence to first order in the magnetic field projection onto the electric field axis. This frequency difference therefore allows direct calibration of the electric field magnitude, by taking into account linear and quadratic DC Stark shifts. 

\subsection{Magnetic field}

Each magnetic field amplitude $B_0$ was determined by performing spectroscopy of the intercombination line at 689 nm between the $5s^2\ ^1S_0$ ground state and the $5s5p\ ^3P_1, m_J = 1$ excited state, using Land\'e $g$-factor $g_{^3P_1} = 1.50116$ obtained from the Russell-Saunders formula. This value is expected to be accurate up to corrections at the $10^{-4}$ level due to multielectron effects \cite{2025Pucher}. The resonant frequency is determined with an uncertainty of 2 kHz, which translates into a magnetic field uncertainty of 1 mG.

In the presence of a residual angle $\theta$ between the directions of the magnetic and electric field, Eq.~\ref{eq:nu_spin} becomes \begin{equation}
\label{eq:nu_spincos}
\nu_{\mathrm{spin}}(n) = \nu_{\mathrm{so}} (n) \cos\theta +  \delta g \frac{\mu_B } h B_0 . 
\end{equation}
In order to measure $\theta$, we measure $\nu_{\mathrm{spin}}(51)$ as we rotate the orientation of the electric field for $B_0=3.160(1)$~G. We estimate the angle between the electric and magnetic fields to be $|\theta|=3(1)^\circ$, corresponding to $\cos\theta = 0.998(1)$. In the experiment, for each value of $B_0$ reported in Fig.\,\ref{fig3}, we cancel the residual magnetic field components along the transverse $(x,y)$ direction using the auxiliary coils axes, ensuring that $\mathbf{B}_0 ={B}_0\hat{\boldsymbol{z}}$. We estimate the relative amplitude of the residual transverse fields after cancelation $B_x / B_0$, $B_y/ B_0$ to be lower than $ 1\%$, corresponding to a variation of the direction of the magnetic field below $1^\circ$.  This results in a relative uncertainty of $10^{-3}$ in the value of $\cos\theta$. In principle, this results in an additional uncertainty on the determination of $\nu_{\mathrm{so}} (n)$ and $\delta g$. However, it is negligible as compared to the relative uncertainty of $10^{-2}$ resulting from the precision of the fit of $\nu_{\mathrm{spin}}(n)$.

\section{Preparation and detection of circular states}
\label{appc}
\subsection{Adiabatic rapid passage}

After three-photon laser excitation to the $5s\,52f\ ^1F_3,\, m_J = m_\ell = -3$ state, we ramp the electric field from zero to $F = 2.22(2)\,\mathrm{V/cm}$ in $1\,\textgreek{m}\mathrm{s}$. A radio-frequency rapid adiabatic passage is then performed. We ramp up a $\sigma_-$-polarized radio-frequency field at $250\,\mathrm{MHz}$ to an amplitude $F_{\mathrm{RF},\sigma_-} = 0.8(2)\,\mathrm{V/m}$ in $1\,\textgreek{m}\mathrm{s}$, then increase the static electric field from $F = 2.22(2)\,\mathrm{V/cm}$ to $2.64(2)\,\mathrm{V/cm}$ in $7\,\textgreek{m}\mathrm{s}$. The radio-frequency field is finally turned off in $1\,\textgreek{m}\mathrm{s}$, leaving atoms in high-angular-momentum states close to the circular state $ m_\ell = -51$. Over the 9 $\mu$s duration of the circularization process, spin-orbit coupling can be neglected so that we prepare circular atoms in a singlet state.

To detect the circular states, we ionize the atoms by ramping the static electric field up to $163(2)\,\mathrm{V/cm}$ in 300 $\,\textgreek{m}\mathrm{s}$ and accelerate the resulting Sr$^+$ ions onto a channeltron. Fig.\,\ref{figMethods_Channeltron}.c shows the ion count as a function of the ionization field.  
The blue points correspond to the ionization signal at the end of the circularization sequence. The populations of the circular states and high-angular momentum states with $ |m_\ell| \lesssim (n-1)$ are not resolved by the ionization detection. To measure the efficiency of circular state preparation, we selectively transfer the $52c$ atoms into $50c$ using the microwave probe method described in \cite{signoles_coherent_2017} (black points). By comparing the area of the black curve in the grey detection window of Fig.\,\ref{figMethods_Channeltron}.c with the total area of the blue data, and taking into account the relative detection efficiencies of the different circular states (see below), we estimate that $66(2)\%$ of the high-angular-momentum Rydberg states are in the $52c$ state at the end of the circularization process.

\subsection{Purification}

To prepare a pure circular state in $51c$, we first transfer the atoms from $52c$ to $51c$ using a microwave pulse. We then implement a cleaning sequence (Fig.\,\ref{figMethods_Channeltron}.b) that removes non-circular atoms left in the 52 multiplicity. This sequence consists in applying two short radio-frequency pulses of duration $60\,\mathrm{ns}$, separated by $28\,\mathrm{ns}$, at the same frequency and with the same amplitude as for the adiabatic passage, at a field $F=2.56(2)\,\mathrm{V/cm}$. This DC electric field value and the RF pulse sequence are optimized to drive the remaining high-angular-momentum $n=52$ states back to low-angular-momentum states, while leaving the $51c$ state unaffected \cite{facon_sensitive_2016}. As the ionization threshold of low-angular-momentum states is much lower than that of the $51c$ state, we apply a short ionization pulse that removes all non-circular atoms, without affecting the $51c$ atoms. The ionization signal of purified $51c$ atoms corresponds to the red points in Fig.\,\ref{figMethods_Channeltron}.c.

To prepare circular states with other principal quantum number $n$ (section \ref{seq-varyn}), we first prepare pure $51c$ atoms as described above, then transfer the atoms to the circular state of interest $nc$ by a second microwave pulse resonant with the $51c \rightarrow nc$ transition. To detect the population of $nc$, we apply at the end of $t_\mathrm{wait}$ a microwave pulse that send back $nc$ into $52c$ (unless $n=52$, in which case we apply no microwave pulse) before transferring back the $52c$ state into the ground state with the sequence described in section \ref{seq-ST} [Fig.\,\ref{fig3}(a)].

\subsection{Circular state decay}

To characterize the decay of the $51c$ level, we measure the population of the ($50c$, $51c$, $52c$) circular states using the colored detection windows defined in Fig.\,\ref{figMethods_Channeltron}.b. To determine the relative detection efficiency between the three different levels, we induce microwave Rabi oscillations from $51c$ to $52c$ or $50c$. We determine correction factors to the populations detected in the $52c$ and $50c$ windows so that the total corrected population is constant during the Rabi oscillation.

To extract a black-body radiation temperature, we model the population dynamics assuming that radiative transfers are dominated by transitions between adjacent circular states $nc \leftrightarrow (n\pm1)c$, thereby neglecting thermal transfer to neighboring elliptical states. The rate-equation model is restricted to the circular ladder spanning states within $\pm 10$ around $n=51$, which is sufficient to capture all relevant population flow during the experimental timescale. The measured signals are fitted using this truncated ladder, while only the experimentally defined detection windows (centered on $50c$, $51c$, and $52c$) are used as observables. This allows us to extract the effective transition rates out of the $51c$ state, from which the lifetime is obtained by summing the outgoing rates.

\section{Singlet-triplet oscillations}
\label{appd}
\subsection{Model for singlet-triplet oscillations}

The electron spin dynamics is  governed by a sum of two Hamiltonians $\hat H = \hat H_c+\hat H_R$, acting independently on the core and Rydberg electrons, with $\hat H_i  = g_i \mu_B \mathbf{B}_i \cdot \hat{\mathbf s}_{i}$, where $ \mu_B$ is the Bohr magneton, $g_{i}$ and $\hat{\mathbf s}_{i}$ the Land\'e $g$-factor and the spin operator, $\mathbf{B}_c=B_0\hat{\boldsymbol{z}}+\mathbf B_\mathrm{eff}- \mathbf B_{\mathrm{so},n}/2$ and $\mathbf{B}_R=B_0\hat{\boldsymbol{z}}+\mathbf B_{\mathrm{so},n}$. Since $|| \mathbf B_0 ||\gg ||\mathbf B_\mathrm{eff}||,||\mathbf B_{\mathrm{so},n}||$, $\ketu \ _c\ketd\ _R$ and $ \ketd \ _c\ketu\ _R$ remains the eigenstates and we have
\begin{eqnarray*}
\label{eq:delta_E}
h \nu_{spin} &= &( g_c ||\mathbf{B}_c|| - g_R ||\mathbf{B}_R|| )\mu_B\\ & \approx& g_c \mu_B\mathbf B_\mathrm{eff}\cdot \hat{\boldsymbol{z}} \\ & & + \ (g_c+g_R/2)\mu_B  \mathbf B_{\mathrm{so},n}\cdot \hat{\boldsymbol{z}}\\ & & + \ \delta g\mu_BB_0
\end{eqnarray*}
to first order in $||\mathbf B_\mathrm{eff}||,||\mathbf B_{\mathrm{so},n}||$.

\subsection{Tweezer effective magnetic field $\mathbf B_\mathrm{eff}$}

In Fig.\,\ref{fig2}.d, the ellipticity angle of the tweezer polarization is extracted from the dependence of the singlet-triplet oscillation frequency on the tweezer power and the bias magnetic field $B_y$. Assuming the tweezers have an elliptical polarization $\boldsymbol{\epsilon}$ in the plane perpendicular to the tweezer propagation axis $\hat{\textbf{y}}'$ (which may not perfectly coincide with the $y$ axis defined by the auxiliary coils wraped around the aspheric lenses), one can parametrize
\begin{equation}
\boldsymbol{\epsilon} = \cos(\gamma)\hat{\textbf{z}}' + \mathrm{i}\sin(\gamma) \hat{\textbf{x}}',
\end{equation}
where $\gamma$ is the ellipticity angle and $\hat{\textbf{z}}', \hat{\textbf{x}}'$ are the axis of the ellipse. This induces a vector light shift of the ionic core, which can be described as an an effective magnetic field whose amplitude scales with the local intensity $I$,
\begin{equation}
\mathbf{B}_\mathrm{eff} = \frac{\alpha_v I \sin\left(2\gamma\right)}{g_c \mu_B \epsilon_0 c} \hat{\mathbf{y}}',
\end{equation}
where $\alpha_v$ is the vector polarizability of the ionic core \cite{2025Steck}. %
The slopes deduced from the linear fit of Fig.\,\ref{fig2}.c are proportional to the projection of $\mathbf{B}_\mathrm{eff} $ along the direction $B_0\hat{\boldsymbol{z}}+B_y\hat{\boldsymbol{y}}$.
Measuring these slopes for several values of $B_y$, we deduce
\begin{equation}
\frac{\alpha_v \sin\left(2\gamma\right)}{g_c \mu_B \epsilon_0 c} = 2.25(5)~\mathrm{mG}\cdot\mathrm{mW}^{-1}\cdot\textgreek{m}\mathrm{m}^2,
\end{equation}
From online atomic data for $^{88}$Sr$^+$ \cite{2025Portal}, we estimate $\alpha_v = -9.8(1)$\,a.u. at wavelength $532$\,nm, which allows us to extract $\gamma = 2.2(1)^\circ$.

\subsection{Spin-orbit coupling}

The magnetic field created by the Rydberg electron motion onto the ionic core is 
\begin{equation}
\label{eq:SO}
\mathbf B_{\mathrm{so},n} =-\frac{e\mu_0}{4\pi m_e} \left\langle\frac{\hat{\mathbf L}}{ \hat r^3}\right\rangle_{\ket{nc}},
\end{equation}
where $e$ is the electron charge, $m_e$ the electron mass, $\mu_0$ the vacuum magnetic permeability, $\hat{\mathbf L}$ the angular momentum operator and $\hat r$ the electron position operator.
In the circular state $m_\ell  <0$, we have $ \left\langle \hat{\mathbf L} \right\rangle_{\ket{nc}} = -(n-1)\hbar \hat{\boldsymbol{u}}_F$, with $\hat{\boldsymbol{u}}_F$ the unit vector parallel to the electric field $\textbf{F}$ and 
\begin{equation} 
\label{eqn-nuso} 
\mathbf B_{\mathrm{so},n} = \frac{\mu_0 e \hbar}{4\pi m_e a_0^3} \frac{1}{n^4(n-1/2)}\hat{\boldsymbol{u}}_F,
\end{equation} 
where  $a_0$ is the Bohr radius and the radial integral $\left\langle\frac{1}{ \hat r^3}\right\rangle_{\ket{nc}}$  is taken from \cite{2003BransdenJoachain}. From this expression, we get $\nu_{\mathrm{so}}(n) $ (green line in Fig.\, \ref{fig4}.a).

\subsection{Effect of magnetic dipole-dipole interactions}

In this paragraph, we estimate the effect of magnetic dipole-dipole interactions between the two electron spins, and show that it can be neglected in the analysis of our experimental results. The corresponding hamiltonian
\begin{equation}
\hat{H}_{\mathrm{mddi}} = \frac{\mu_0}{4\pi \hat{r}^3} \frac{g_cg_R \mu_B^2}{\hbar^2} \left[\hat{\mathbf{s}}_c \cdot \hat{\mathbf{s}}_R - 3(\hat{\mathbf{s}}_c \cdot \hat{\mathbf{r}})(\hat{\mathbf{s}}_R \cdot \hat{\mathbf{r}})\right]
\end{equation}
is symmetric upon exchange of the two spins, and therefore does not couple the singlet and triplet states, preserving the symmetric or antisymmetric character of any spin state. It does, however, shift the energies of the singlet $|S\rangle \equiv \frac{1}{\sqrt 2}( \ketu \ _c\ketd\ _R - \ketd \ _c\ketu\ _R )$ and triplet $|T_0\rangle \equiv \frac{1}{\sqrt 2}( \ketu \ _c\ketd\ _R + \ketd \ _c\ketu\ _R )$  states differently, with
\begin{equation}
\label{eq:mddi}
\langle nc, T_0 | \hat{H}_{\mathrm{mddi}} | nc, T_0 \rangle =  \frac{\mu_0g_cg_R \mu_B^2}{4\pi}  \left\langle\frac{1}{ \hat r^3}\right\rangle_{\ket{nc}},
\end{equation}
and \begin{equation}\langle  nc, S | \hat{H}_{\mathrm{mddi}} | nc, S \rangle  = 0.\end{equation} This corresponds to an energy shift of $20$\,Hz, smaller by a factor $1/n$ compared to the spin-orbit coupling considered in the main text. Its contribution to the singlet-triplet oscillation frequency will only appear at second order and can therefore be neglected.

\end{document}